\documentclass{aastex631}
\usepackage{newtxtext,newtxmath}
\shorttitle{Strong depletion of $^{13}$C in CO induced by photolysis of CO$_{2}$ in the Martian atmosphere}
\shortauthors{Yoshida et al.}
\graphicspath{{./}{figures/}}

\begin{document}

\title{Strong depletion of $^{13}$C in CO induced by photolysis of CO$_{2}$ in the Martian atmosphere calculated by a photochemical model}

\author{Tatsuya Yoshida}
\affiliation{Department of Geophysics, Graduate School of Science, Tohoku University, Sendai, Miyagi, Japan}

\author{Shohei Aoki}
\affiliation{Department of Complexity Science and Engineering, Graduate School of Frontier Sciences, University of Tokyo, Kashiwa, Chiba, Japan}

\author{Yuichiro Ueno}
\affiliation{Department of Earth and Planetary Sciences, Tokyo Institute of Technology, Meguro, Tokyo, Japan}
\affiliation{Earth-Life Science Institute (WPI-ELSI), Tokyo Institute of Technology, Meguro, Tokyo, Japan}
\affiliation{Japan Agency for Marine-Earth Science and Technology (JAMSTEC), Yokosuka, Kanagawa, Japan}

\author{Naoki Terada}
\affiliation{Department of Geophysics, Graduate School of Science, Tohoku University, Sendai, Miyagi, Japan}

\author{Yuki Nakamura}
\affiliation{Department of Geophysics, Graduate School of Science, Tohoku University, Sendai, Miyagi, Japan}

\author{Kimie Shiobara}
\affiliation{Department of Geophysics, Graduate School of Science, Tohoku University, Sendai, Miyagi, Japan}

\author{Nao Yoshida}
\affiliation{Department of Geophysics, Graduate School of Science, Tohoku University, Sendai, Miyagi, Japan}

\author{Hiromu Nakagawa}
\affiliation{Department of Geophysics, Graduate School of Science, Tohoku University, Sendai, Miyagi, Japan}

\author{Shotaro Sakai}
\affiliation{Department of Geophysics, Graduate School of Science, Tohoku University, Sendai, Miyagi, Japan}

\author{Shungo Koyama}
\affiliation{Department of Geophysics, Graduate School of Science, Tohoku University, Sendai, Miyagi, Japan}



\begin{abstract}
	The isotopic signature of atmospheric carbon offers a unique tracer for the history of the Martian atmosphere and the origin of organic matter on Mars. Photolysis of CO$_{2}$ is known to induce strong isotopic fractionation of carbon between CO$_{2}$ and CO. However, its effect on the carbon isotopic compositions in the Martian atmosphere remains uncertain. Here we develop a 1-D photochemical model considering isotopic fractionation via photolysis of CO$_{2}$ to estimate the vertical profiles of the carbon isotopic compositions of CO and CO$_{2}$ in the Martian atmosphere. We find that CO is depleted in $^{13}$C compared with CO$_{2}$ at each altitude due to the fractionation via CO$_{2}$ photolysis: the minimum value of $\delta ^{13}$C in CO is about $-170$ ‰ under the standard eddy diffusion setting. This result supports the hypothesis that fractionated atmospheric CO is responsible for the production of the $^{13}$C-depleted organic carbon in Martian sediments detected by Curiosity Rover through the conversion of CO into organic materials and their deposition on the surface. The photolysis and transport-induced fractionation of CO we report here leads to a $\sim 15$ \% decrease in the amount of inferred atmospheric loss when combined with the present-day fractionation of the atmosphere and previous studies of carbon escape to space. The fractionated isotopic composition of CO in the Martian atmosphere may be observed by ExoMars Trace Gas Orbiter (TGO) and ground-based telescopes, and escaping ion species produced by the fractionated carbon-bearing species may be detected by Martian Moons eXploration (MMX) in the future. 
\end{abstract}



\section{Introduction} \label{sec:intro}
	Isotopic compositions of volatiles have been used to trace the history of planetary atmospheres. The enrichment in the heavy isotopes of the atmospheric components such as hydrogen, carbon, nitrogen, and noble gases of Mars with respect to Earth and primitive meteorites indicates that Mars has lost a large portion of the atmosphere by atmospheric escape processes (e.g., Owen, 1977; Jakosky, 1991; Jakosky et al., 1994; Pepin, 1991; 1994; Hu et al., 2015; Kurokawa et al., 2018). 
	
	The isotopic signature of carbon-bearing species offers a unique tracer for the atmospheric evolution of Mars since CO$_{2}$ is the major constituent of the Martian atmosphere (Hu et al., 2015). Hu et al. (2015) modeled the isotopic fractionation of carbon induced by atmospheric escape processes such as photochemical escape and solar-wind-induced sputtering, deposition of carbonate minerals, and volcanic outgassing to trace the evolution of the carbon reservoir and its isotopic composition to satisfy the present-day carbon isotopic ratio of CO$_{2}$ in the atmosphere observed by the Curiosity Rover. In their calculation, atmospheric escape, especially photochemical escape via CO photodissociation, enriches heavy carbon ($^{13}$C) in the atmosphere efficiently, which can drive the carbon isotopic ratio to the present-day fractionated value.
	
	In addition to the isotopic fractionation processes considered by Hu et al. (2015), photolysis of CO$_{2}$ is expected to affect the isotopic compositions of carbon-bearing species significantly. Schmidt et al. (2013) demonstrated that the UV absorption cross-section of $^{13}$CO$_{2}$ is several hundred per mil lower than that of $^{12}$CO$_{2}$ in the wavelength range of 138-212 nm by quantum mechanical methodology. This suggests that the photolysis of CO$_{2}$ could induce isotopic fractionation between CO$_{2}$ and carbon-bearing photochemical products such as CO in the troposphere and stratosphere by several hundred per mil. The degree of the fractionation is much higher than those of other known isotopic fractionation processes. For example, condensation of carbonate minerals, one of the other fractionation processes of carbon, enriches the carbon isotopic ratio of carbonate precipitates only by $\sim 10$ ‰ relative to the source atmosphere (Hu et al., 2015). However, the effect of photo-induced isotopic fractionation on the carbon isotopic composition in the Martian atmosphere has not been investigated quantitatively.
	
	The photo-induced carbon isotopic fractionation in CO may be related to the carbon isotopic composition of organic carbon in Martian sediments. It was predicted that $^{13}$C-depleted organic materials can have been deposited on the surface if their photochemical production via CO as an intermediate proceeded efficiently on early Mars (Lammer et al., 2020; Stueken et al., 2020). As expected, Curiosity Rover found that $\sim 3.5$ billion years old sedimentary organic carbon at Gale crater is depleted in $^{13}$C by more than $\sim 100$ ‰ compared with that of the atmosphere (House et al., 2022). In response to this, Ueno et al. (2022) has suggested that atmospheric synthesis of organic materials from CO is a plausible mechanism to explain the presence of organic carbon in early Martian sediments and its strong $^{13}$C depletion through the experimental and theoretical studies of photolysis of CO$_{2}$. To validate this hypothesis, quantitative estimates of the carbon isotopic composition in CO considering chemical kinetics and transport in the atmosphere are needed.
	
	The photo-induced carbon isotopic fractionation in CO may also affect the degree of isotopic fractionation by atmospheric escape. Photodissociation of CO is the most important photochemical source of escaping carbon atoms from Mars (Fox and Bakalian, 2001; Groeller et al., 2014; Lo et al., 2021). Here, considering vertical transport of the fractionated CO to the upper atmosphere near the escape region should lead to a change in the fractionation factor of photochemical escape via CO photodissociation.

	In this study, we develop a 1-D atmospheric photochemical model considering the isotopic fractionation via photolysis of CO$_{2}$ in order to quantitatively estimate the vertical profiles of the carbon isotopic compositions of CO and CO$_{2}$. It allows us to clarify the effect of the photolysis on the isotopic compositions. This paper is organized as follows. In Section 2, we describe the outline of our 1-D photochemical model. In Section 3, we show the numerical results of the atmospheric profiles. In Section 4.1, we discuss the dependence of the isotopic composition profiles on the magnitude of the eddy diffusion coefficient. In Section 4.2, we discuss the relationship between the fractionated atmospheric CO and $^{13}$C-depleted organic carbon in Martian sediments. In Section 4.3, we discuss the effect of the fractionation by the photolysis on the degree of the fractionation through atmospheric escape. In Section 4.4, the detectability of the calculated isotopic fractionation in CO by the existing measurements is discussed.

\section{Model description} \label{sec:model}
	We use a 1-D photochemical model developed by Nakamura et al. (2022b, c) with some modifications to the chemical processes. It solves continuity-transport equations that govern changes in the number density profiles of each chemical species by numerical integration over time until the profiles settle into steady states. As for the chemical processes, 57 chemical reactions are considered for 17 species: $^{12}$CO$_{2}$, $^{13}$CO$_{2}$, $^{12}$CO, $^{13}$CO, H$_{2}$O, O, O($^{1}$D), H, OH, H$_{2}$, O$_{3}$, O$_{2}$, HO$_{2}$, H$_{2}$O$_{2}$, HO$^{12}$CO, HO$^{13}$CO, and $^{12}$CO$_{2}^{+}$ (Table A1). Here we refer to the chemical species and reactions considered by Chaffin et al. (2017). We newly include minor carbon-bearing isotopologues such as $^{13}$CO$_{2}$, $^{13}$CO, and HO$^{13}$CO and their chemical reactions. To calculate the profiles of photolysis rates, we adopt the solar spectrum profile in the wavelength from 0.5 to 1100 nm obtained by Woods et al. (2009) and solve the radiative transfer by considering the absorption of the solar irradiation by chemical species. We adopt the absorption cross-section of $^{12}$CO$_{2}$ and $^{13}$CO$_{2}$ at 138-212 nm provided by Schmidt et al. (2013) to estimate the isotopic fractionation through the photolysis. For the absorption cross-sections of $^{12}$CO$_{2}$ and $^{13}$CO$_{2}$ in other wavelengths, we refer to Huestis and Berkowitz (2011) and references therein. For the absorption cross-sections of the other chemical species, we mainly refer to the JPL publication (Burkholder et al., 2015) and the MPI-Mainz-UV-VIS Spectral Atlas of Gaseous Molecules (Keller-Rudek et al., 2013; \url{https://www.uv-vis-spectral-atlas-mainz.org/uvvis/}). The spectral bin at 138-212 nm is 0.1 nm to resolve the difference in the photolysis rate between $^{12}$CO$_{2}$ and $^{13}$CO$_{2}$, and that at the other wavelengths is 1 nm. We also consider the difference in the rate coefficients of the chemical reactions between $^{12}$CO and $^{13}$CO: the rate coefficient of the reaction of $^{13}$CO with O (R22 in Table A1) is 1.0074 times as large as that of $^{12}$CO (R21) (Ueno et al., 2022) and the rate coefficients of the reactions of $^{13}$CO with OH (R52 and R54) is 0.9891 times as large as those of $^{12}$CO (R51 and R53) (Feilberg et al., 2005). In addition, the rate coefficient of the reaction of HO$^{13}$CO with O$_{2}$ (R56) is assumed to be 0.9891 times as large as that of HO$^{12}$CO (R55) by referring to the difference in the rate coefficient between R51(R53) and R52(R54). We adopt the fixed number density profile of H$_{2}$O used by Koyama et al. (2021). Here the relative humidity below 30 km is fixed at 22 \% to give 9.5 precipitable microns of water and the H$_{2}$O profile above is connected to saturation water vapor at the altitude where the temperature is minimum; then, the same mixing ratio is assumed above higher altitudes. The number density profile of $^{12}$CO$_{2}^{+}$ is fixed as the standard case of Chaffin et al. (2017). The profiles of temperature, eddy diffusion coefficient, and binary diffusion coefficient are taken from Chaffin et al. (2017).
	
	The lower boundary is set at the planetary surface. The number densities of $^{12}$CO$_{2}$ and $^{13}$CO$_{2}$ at the lower boundary are fixed at $2.10\times 10^{17}\,\mathrm{cm^{-3}}$ (Chaffin et al., 2017) and $2.46\times 10^{15}\,\mathrm{cm^{-3}}$ respectively to satisfy the carbon isotopic ratio in CO$_{2}$ measured by the Sample Analysis at Mars' Tunable Laser Spectrometer (SAM/TLS) on the Curiosity Rover (Webster et al., 2013). The altitude of the upper boundary is set at 200 km. As the upper boundary condition, H and H$_{2}$ are assumed to escape to space by Jeans escape, and the O escape rate is fixed at $1.2\times 10^{8}\,\mathrm{cm^{-2}\,s^{-1}}$ as Chaffin et al. (2017). 
	
\begin{figure}[htbp]
	\centering
	\includegraphics[width=0.45\columnwidth]{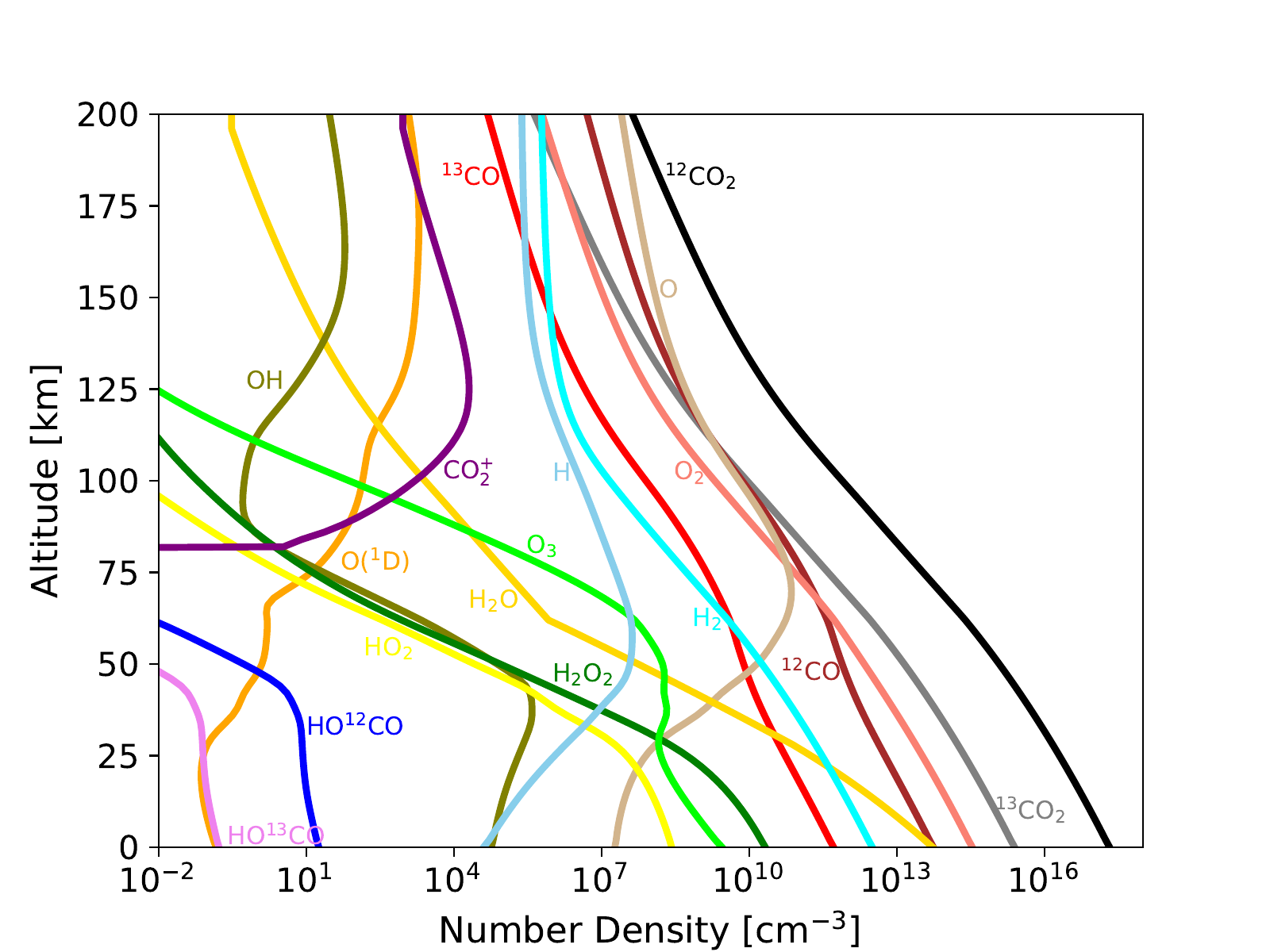}
	\caption{Number density profiles of each chemical species.}
\end{figure}
	
\section{Results} \label{sec:result}
	The number density profiles of each chemical species in the steady state are shown in Fig. 1. The calculated profiles of major isotopologues are in good agreement with Chaffin et al. (2017) except for the slight difference in the abundances of odd hydrogen and odd oxygen caused by the difference in the adopted absorption cross-sections and H$_{2}$O profile. Fig. 2 represents the profiles of the carbon isotopic ratios in CO and CO$_{2}$. Here the carbon isotopic ratios are expressed by the deviation of the calculated ratio with respect to the standard ratio in units per mil:
	\begin{equation}
		\delta ^{13}\mathrm{C}=\left(\frac{R}{R_{s}}-1\right)\times 1000
	\end{equation}
	where $R$ is the $^{13}$C/$^{12}$C ratio and $R_{s}=1.123\times 10^{-2}$ which is the $^{13}$C/$^{12}$C ratio of the Vienna Pee Dee Belemnite (VPDB). As shown in Fig. 2, CO has lower $\delta ^{13}$C than CO$_{2}$ at each altitude because photolysis of CO$_{2}$, which is the main formation reaction of CO, involves the isotopic fractionation of carbon between CO$_{2}$ and CO due to the difference in the absorption cross-section between $^{12}$CO$_{2}$ and $^{13}$CO$_{2}$ (Shmidt et al., 2013; Fig. 3).
	
	Below $\sim 100$ km, $\delta ^{13}$C in CO decreases as the altitude decreases: it takes the minimum value of about $-170$ ‰ near the surface. The degree of the isotopic fractionation is much higher than those of other fractionation processes such as condensation of carbonate minerals (Hu et al., 2015). The reason why $\delta ^{13}$C in CO decreases with decreasing altitude is that the wavelength of the absorbed solar irradiation is longer in the lower region (Fig. 3(a)), where the difference in the absorption cross-section between $^{12}$CO$_{2}$ and $^{13}$CO$_{2}$ is large (Fig. 3(b),(c)).
	
	The carbon isotopic ratio in CO$_{2}$ below $\sim 100$ km is constant at the surface value of $46$ ‰ which is assumed to be equal to the value measured by SAM/TLS on Curiosity Rover. On the other hand, the average $\delta ^{13}$C in the altitude range of $70-90$ km measured by the Atmospheric Chemistry Suite (ACS) onboard TGO is $-3 \pm 37$ ‰ (Alday et al., 2021b). The reason for the difference in the evaluated carbon isotopic ratio in CO$_{2}$ between our model (SAM/TLS) and ACS remains uncertain. Alday et al. (2021b) suggested two scenarios to reconcile both measurements by SAM/TLS and ACS as follows. One possible scenario requires the presence of unknown isotopic fractionation processes between the lower and upper atmospheres of Mars: ACS measures the carbon isotopic ratio above 70 km while the Curiosity Rover measures that on the surface. The other relies on the impact of climatological isotopic fractionation: the ACS measurements extend over a large range of locations, seasons, and local times, which allows averaging over hundreds of measurements, from which the effects of climatological fractionation are expected to be small although the measurements made by the Curiosity Rover are always made in the same location, at roughly the same local time.
	
	$\delta ^{13}$C in both CO and CO$_{2}$ decreases as the altitude increases above $\sim 100$ km, which corresponds to the altitude of the homopause, by the diffusive separation resulting from the difference in the molecular mass between isotopologues. The calculated isotopic fractionation of CO$_{2}$ in the upper region above $\sim 100$ km is consistent with the profile observed by ACS (Alday et al., 2021b). 
	
	The dotted line in Fig. 2 represents the ratio of the photolysis rate of $^{13}$CO$_{2}$ to that of $^{12}$CO$_{2}$, which approximates the isotopic ratio of CO when assuming the local photochemical equilibrium without vertical transport. The difference between the dotted line and the solid orange line shows the effect of the vertical transport by diffusion on the profile of the CO isotopic composition: the vertical transport dilutes the difference in the isotopic composition among altitudes.

\begin{figure}[htbp]
	\centering
	\includegraphics[width=0.45\columnwidth]{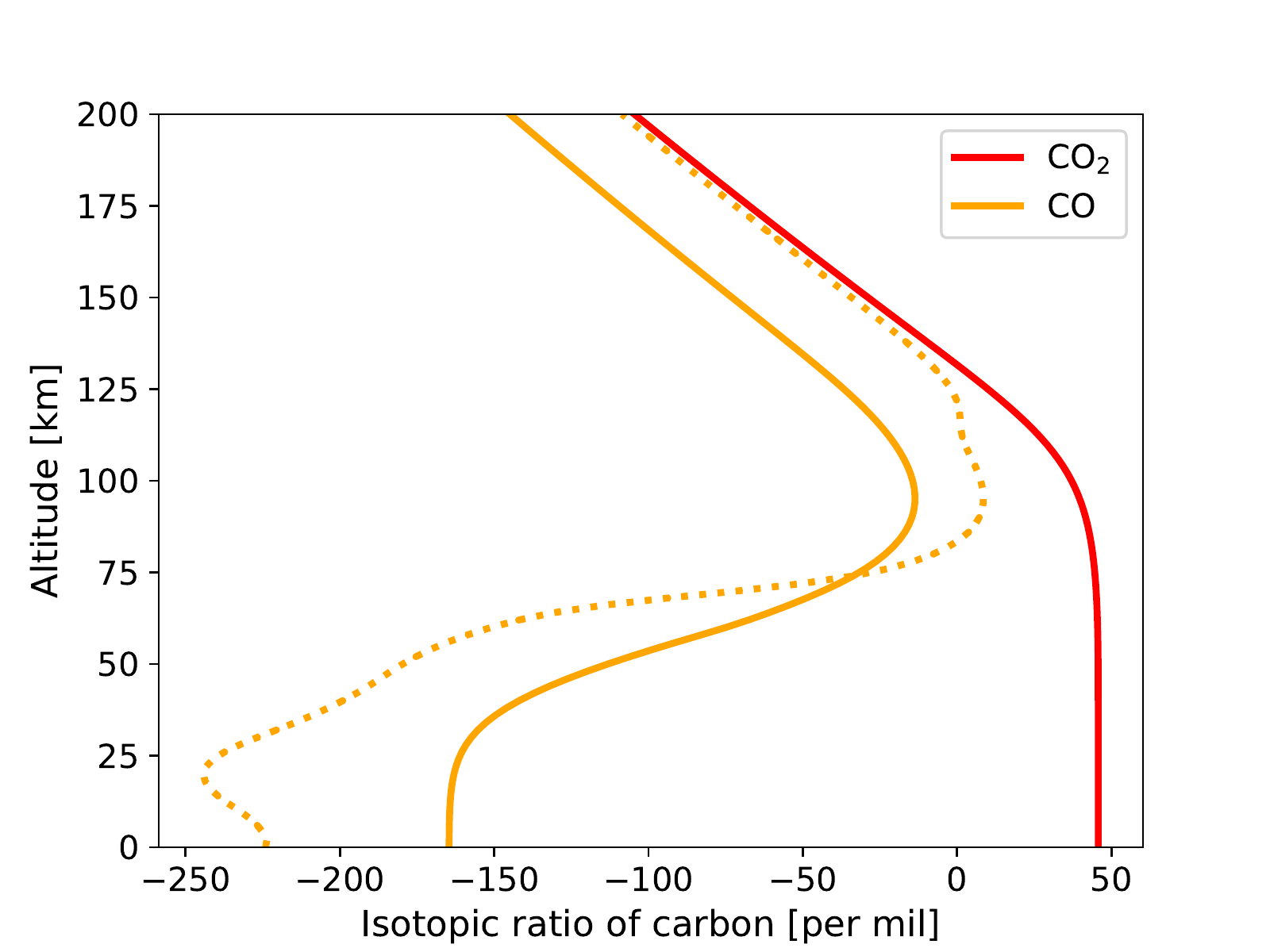}
	\caption{Profiles of the carbon isotopic ratios in CO and CO$_{2}$. The isotopic ratios are expressed by the deviation of the calculated ratio with respect to the standard ratio in units per mil: $\delta ^{13}\mathrm{C}=\left(\frac{R}{R_{s}}-1\right)\times 1000$, where $R$ is the $^{13}$C/$^{12}$C ratio and $R_{s}=1.123\times 10^{-2}$. The orange solid line and red solid line represent the carbon isotopic ratio in CO and CO$_{2}$, respectively. The dotted orange line represents the ratio of the photolysis rate of $^{13}$CO$_{2}$ to that of $^{12}$CO$_{2}$, which approximates the isotopic ratio of CO when assuming the local photochemical equilibrium without vertical transport.}
\end{figure}

\begin{figure}[htbp]
	\centering
	\includegraphics[width=0.45\columnwidth]{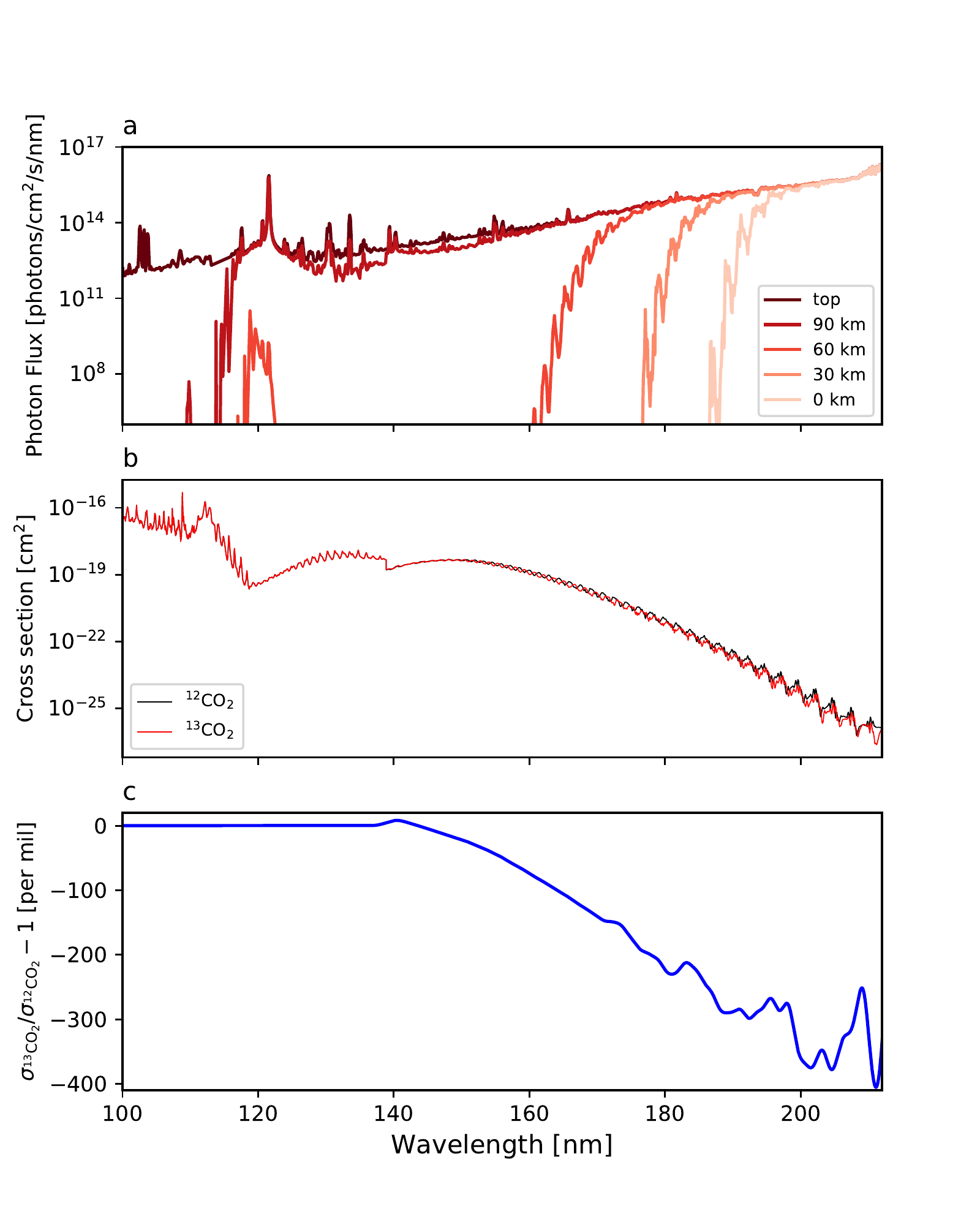}
	\caption{(a) Profiles of photon flux with wavelength at each altitude. (b) Absorption cross-sections of $^{12}$CO$_{2}$ and $^{13}$CO$_{2}$ with wavelength. The black and red lines represent the absorption cross-sections of $^{12}$CO$_{2}$ and $^{13}$CO$_{2}$, respectively. (c) The relative difference of absorption cross-section between $^{12}$CO$_{2}$ and $^{13}$CO$_{2}$: $(\sigma_{\rm ^{13}CO_{2}}$/$\sigma_{\rm ^{12}CO_{2}}-1)\times 1000$ where $\sigma_{\rm ^{12}CO_{2}}$ and $\sigma_{\rm ^{13}CO_{2}}$ are the absorption cross-sections of $^{12}$CO$_{2}$ and $^{13}$CO$_{2}$, respectively. Here it is averaged over a Gaussian window with FWHMs of 2.5 nm as done by Schmidt et al. (2013).}
\end{figure}

\section{Discussion} \label{sec:discussion}
\subsection{Effect of change in the magnitude of eddy diffusion coefficient on atmospheric profile}
	The eddy diffusion coefficient in the Martian atmosphere is estimated to be variable with season and latitude (Yoshida et al., 2022). The change in the eddy diffusion coefficient affects the atmospheric profile including the isotopic composition. Fig.4 compares the profiles of the carbon isotopic ratios of CO and CO$_{2}$ under the eddy diffusion coefficient 10 times as small as the standard setting with those under the standard setting. Below $\sim 25\,\mathrm{km}$, $\delta ^{13}$C in CO becomes lower under the lower eddy diffusion coefficient due to the suppression of the dilution of the difference in the isotopic composition among altitudes by vertical transport. In the upper region above $\sim 75$ km, $\delta ^{13}$C in both CO and CO$_{2}$ becomes lower due to the more efficient diffusive separation through the molecular diffusion relative to the mixing by the eddy diffusion. To the contrary, under higher eddy diffusion coefficient, the dilution of the difference in the isotopic composition among altitudes is enhanced, and the diffusive separation through the molecular diffusion is suppressed.
	

\begin{figure*}
	\gridline{\fig{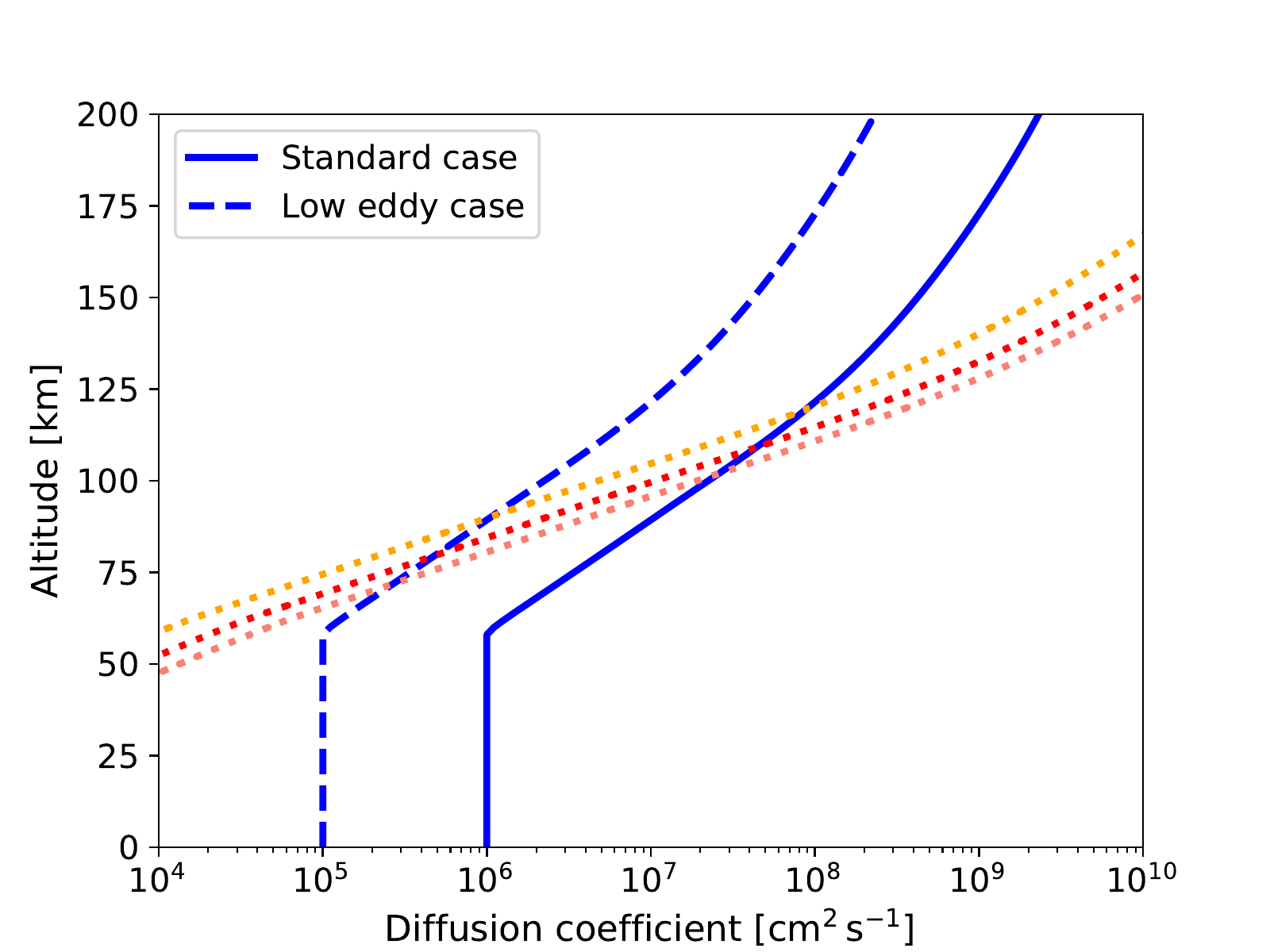}{0.45\textwidth}{(a)}
        	 	 \fig{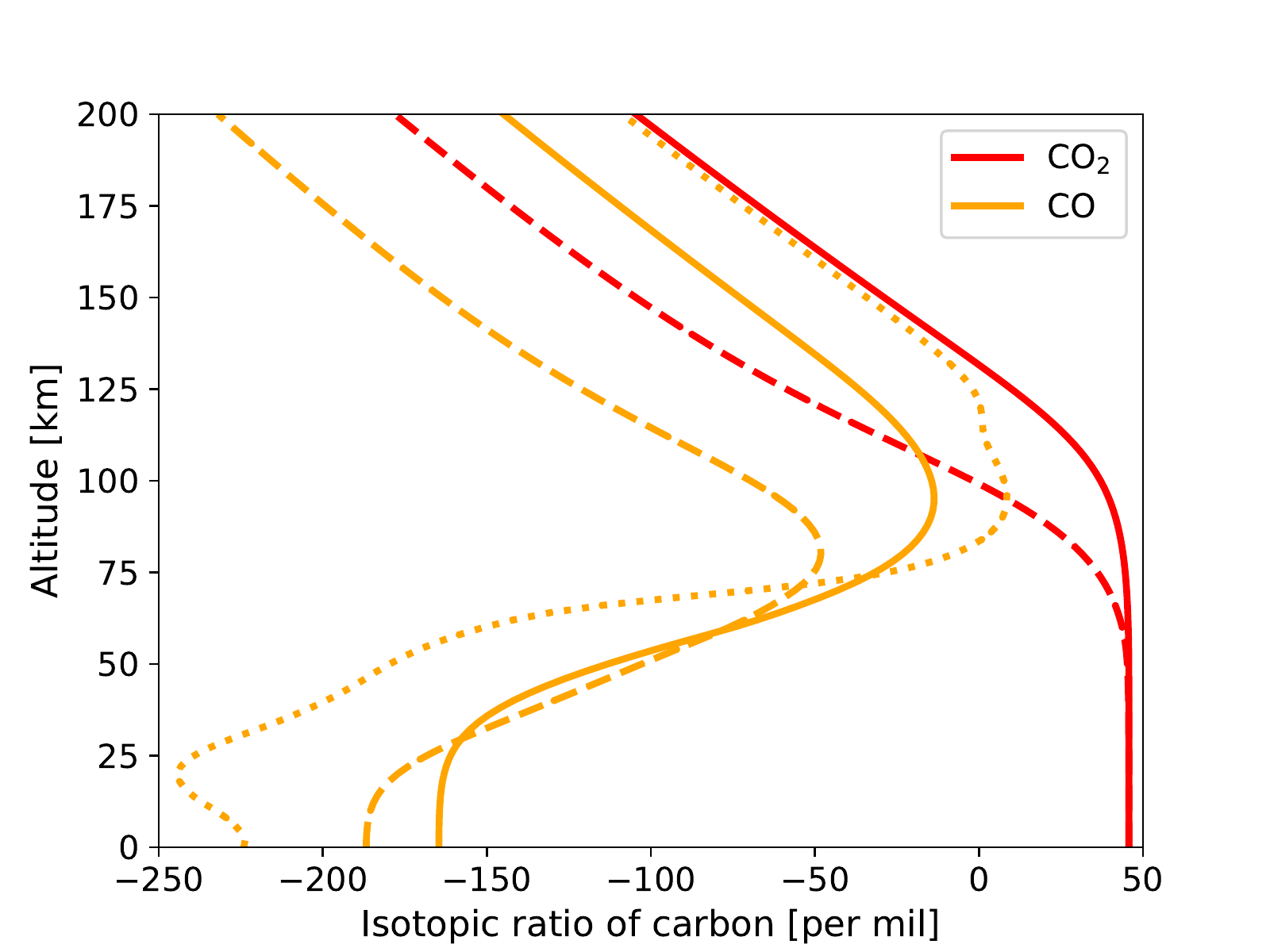}{0.45\textwidth}{(b)}
          }
	\caption{(a) Profiles of the eddy diffusion coefficient and the molecular diffusion coefficient. The solid blue line represents the profile of the eddy diffusion coefficient under the standard setting (``Standard case"), and the dashed blue line represents that 10 times as small as the standard setting (``Low eddy case"). The dotted orange, red, and pink lines represent the profiles of the molecular diffusion coefficients of O, H$_{2}$, and H, respectively. (b) Profiles of the carbon isotopic ratios in CO and CO$_{2}$ depending on the magnitude of the eddy diffusion coefficient. The isotopic ratios are expressed by the deviation of the calculated ratio with respect to the standard ratio in units per mil: $\delta ^{13}\mathrm{C}=\left(\frac{R}{R_{s}}-1\right)\times 1000$, where $R$ is the $^{13}$C/$^{12}$C ratio and $R_{s}=1.123\times 10^{-2}$. The solid lines and dashed lines show the results under the standard eddy diffusion setting and those under the eddy diffusion coefficient 10 times as small as the standard setting, respectively. The orange lines and red lines represent the carbon isotopic ratios in CO and CO$_{2}$, respectively. The dotted orange line is the same as that in Fig. 2, which represents the isotopic ratio of CO on the assumption of the local photochemical equilibrium.
	\label{fig:pyramid}}
\end{figure*}

\subsection{Relationship between the fractionated atmospheric CO and $^{13}$C-depleted organic carbon in Martian sediments}	
	Our results that CO is depleted in $^{13}$C support the hypothesis that photochemical production and deposition of organic materials via CO is responsible for producing the $^{13}$C-depleted organic carbon in Martian sediments (Lammer et al., 2020; Stueken et al., 2020; Ueno et al., 2022). The main organic molecule produced in an early Martian CO$_{2}$-dominated atmosphere should be formaldehyde (H$_{2}$CO), which can be produced as follows (e.g., Pinto et al., 1980):
	\begin{equation}
		\mathrm{H}+\mathrm{CO}+\mathrm{M} \to \mathrm{HCO}+\mathrm{M},
	\end{equation}
	\begin{equation}
		\mathrm{HCO}+\mathrm{HCO}+\mathrm{M}\to \mathrm{H_{2}CO}+\mathrm{CO}+\mathrm{M},
	\end{equation}
	where M is the third body. Considering the production processes, the $^{13}$C-depleted isotopic composition of CO is expected to be transferred to formaldehyde. Assuming that formaldehyde has the same isotopic composition as CO, our results can explain the existence of organic carbon with $\delta ^{13}$C lower than $-100$ ‰ detected by Curiosity Rover. On the other hand, this study does not suppose an early Martian atmosphere condition where organic materials could have been produced. It is our future work to estimate the deposition rate and isotopic composition of organic molecules such as formaldehyde on the condition of early Mars. 

\subsection{Effect of the carbon isotopic fractionation in CO on the fractionation through photochemical escape}
	The carbon isotopic fractionation between the lower atmosphere and the escape region is expected to affect the degree of the fractionation through atmospheric escape. In this section, we estimate the effect roughly by using the framework of Rayleigh fractionation (Rayleigh, 1895; Hunten, 1982). As for atmospheric escape processes, we consider only photochemical escape via CO photodissociation, which is the dominant process for the production of escaping carbon atoms on Mars (e.g., Fox \& Bakalian, 2001; Groeller et al., 2014; Lo et al., 2021). The relationship between the isotopic ratio and the fractionation factor is given by
	\begin{equation}
		\frac{R}{R_{0}}=\left(\frac{N_{^{12}\mathrm{C}}}{N_{^{12}\mathrm{C}}^{0}}\right)^{f-1}
	\end{equation}
	where $R$ and $R_{0}$ are the $^{13}$C/$^{12}$C ratio and its initial value in the whole atmosphere, $f$ is the fractionation factor, $N_{^{12}\mathrm{C}}$ and $N_{^{12}\mathrm{C}}^{0}$ are the total inventory of $^{12}$C and its initial value. Here we define the net fractionation factor as follows:
	\begin{equation}
		f=f_{\mathrm{s-e}}\times f_{\mathrm{esc}}
	\end{equation}
	where $f_{\mathrm{s-e}}$ is the fractionation factor between the surface and the escape region, which is the ratio of $^{13}$C/$^{12}$C in CO at the escape region to that in CO$_{2}$ at the surface, and $f_{\mathrm{esc}}$ is the fractionation factor by the atmospheric escape. We assume that the fractionation factor by the photochemical escape via CO photodissociation $f_{\rm esc}$ is 0.6 referring to Hu et al. (2015) and the altitude of the escape region is 160 km where the production of escaping atoms typically peaks (e.g., Fox and Ha\'{c}, 2009; Lo et al., 2021). The net fractionation factor $f$ under the standard eddy diffusion setting is 0.52. It highly depends on the magnitude of the eddy diffusion coefficient. For example, $f=0.46\,(0.54)$ under the eddy diffusion coefficient 10 times as small (large) as the standard setting.
	
	Fig. 5 shows the $^{13}$C/$^{12}$C ratio relative to the initial value in the whole atmosphere as a function of the fraction of gas lost to space. The carbon fractionation when considering the isotopic fractionation between the lower atmosphere and the escape region proceeds more efficiently than the estimates by Hu et al. (2015). The $^{13}$C/$^{12}$C ratio of the present Martian CO$_{2}$ atmosphere is higher by a factor of about 1.07 than that of mantle-degassed CO$_{2}$ derived from the magmatic component of the SNC meteorites (Hu et al., 2015). Beginning with an atmospheric $^{13}$C/$^{12}$C ratio equal to that of mantle-degassed CO$_{2}$, the effective carbon isotopic fractionation through photochemical escape can drive the carbon isotopic ratio to the present-day fractionated value even when the amount lost by atmospheric escape is small compared with that evaluated by Hu et al. (2015). On the other hand, there are various isotopic fractionation processes such as other atmospheric escape processes like solar-wind-induced sputtering and ion pickup, volcanic outgassing, deposition of carbonate minerals and organic matters, and so on. It is our future work to estimate the evolution of the carbon reservoir and its isotopic composition by considering these processes comprehensively. 
	
	\begin{figure}[htbp]
	\centering
	\includegraphics[width=0.45\columnwidth]{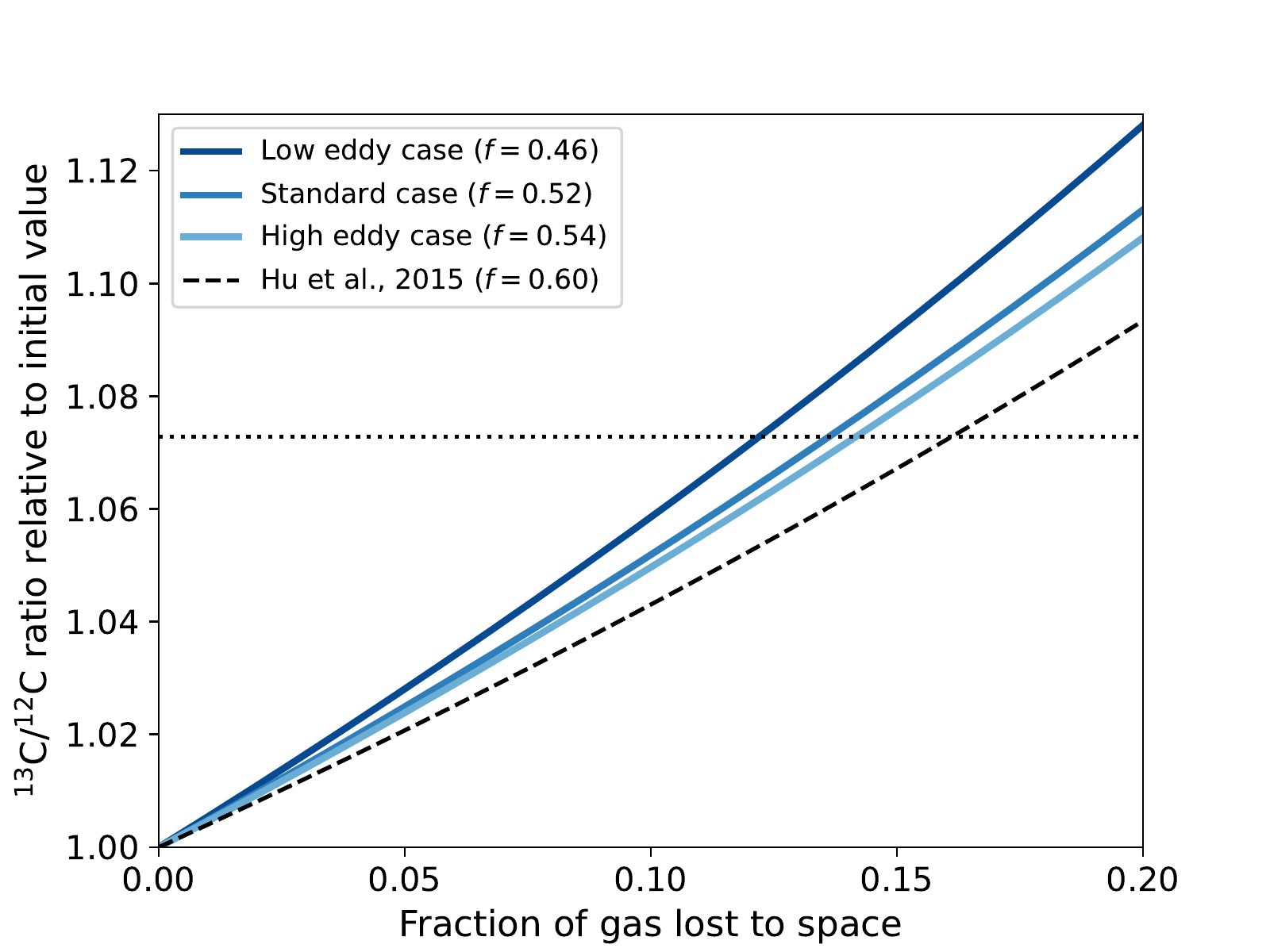}
	\caption{$^{13}$C/$^{12}$C ratio relative to the initial value in the whole atmosphere as a function of the fraction of gas lost to space for each fractionation factor on the standard eddy diffusion coefficient setting ( ``Standard case"), the eddy diffusion coefficient 10 times as small as the standard setting (``Low eddy case"), the eddy diffusion coefficient 10 times as large as the standard setting (``High eddy case"), and the setting of Hu et al. (2015) (``Hu et al., 2015"), respectively. The dotted black line represents the $^{13}$C/$^{12}$C ratio of the present Martian atmosphere relative to that of mantle-degassed CO$_{2}$ derived from the magmatic component of the SNC meteorites (Hu et al., 2015).}
\end{figure}
	
\subsection{Possibility of observing the fractionated carbon isotopic composition}
	Even though a relatively strong depletion of $^{13}$C in CO is suggested by our calculations, the isotopic ratio between $^{12}$CO and $^{13}$CO has not been quantified by previous observations. Strong $^{12}$CO and $^{13}$CO lines are available in the near-infrared spectral range at 4140-4220 cm$^{-1}$, which can be used to measure the isotopic ratio. For that, a high-resolution spectroscopy is required since the lines of $^{12}$CO and $^{13}$CO in the spectral range are quite narrow under the condition of the Mars atmosphere. Spectrometers onboard ExoMars Trace Gas Orbiter (TGO), Nadir and Occultation for Mars Discovery (NOMAD; Vandaele et al., 2018) and Atmospheric Chemistry Suite (ACS; Korablev et al., 2018), are able to perform such a spectroscopic measurements of these spectral ranges with relatively high spectral resolution ($R\sim$16,000-50,000). These instruments perform solar occultation measurements, which makes it possible to achieve high SNR (> 1000) and investigate vertical profiles of trace gas. In fact, these instruments have revealed the vertical profiles of D/H and $^{18}$O/$^{17}$O/$^{16}$O in water vapor (Villanueva et al., 2021, 2022; Alday et al., 2021a) and $^{13}$C/$^{12}$C and $^{18}$O/$^{17}$O/$^{16}$O in CO$_{2}$ (Alday et al., 2021b) and measuring $^{13}$C/$^{12}$C in CO was listed as one of the science targets (Vandaele et al., 2018). Fig. 6(a) shows the expected spectra of NOMAD at 26 km calculated by Asimut radiative transfer code (Vandaele et al., 2006; Aoki et al., 2019). The SNR of the NOMAD spectrum is typically greater than 1000 for a single spectrum, thus the synthetic spectra shown in Fig. 6(a) demonstrate that the strong depletion of $^{13}$C in CO suggested by our calculations can be identified by the NOMAD observations. The other potential platform to measure $^{13}$C/$^{12}$C in CO is a high spectral resolution spectrograph installed at large ground-based telescopes (such as IRTF/iSHELL, VLT/CRIRES+, etc.). They cannot perform solar occultation measurements, however, the spectral resolution of these instruments is about 2-5 times better than that of NOMAD and ACS. Fig. 6(b) shows the expected spectra of the IRTF/iSHELL calculated by PSG radiative transfer code (Villanueva et al., 2018). The SNR of a Mars spectrum taken by IRTF/iSHELL binning over a few pixels (which corresponds to $\sim$1.0”) is typically greater than 100. Given the angular diameter of Mars is greater than 10” in an optimal observation period, the global average of the retrievals can detect the suggested depletion of $^{13}$C in CO with IRTF/iSHELL. Note that the signal from Mars is attenuated by the absorption due to the telluric atmosphere (shown as the black dotted curve in FIg. 6(b)), however the target CO features are not heavily overlapped with the telluric features. Moreover, the telluric features can be removed by modeling them with radiative transfer calculations (e.g., see Villanueva et al., 2013).
	
	Our results suggest that the degree of the isotopic fractionation of escaping carbon is enhanced through vertical transport of the fractionated CO to the upper atmosphere near the escape region (Section 4.3). Recently, the fluxes of C$^{+}$ in the Martian magnetotail have been detected by the Mars Atmospheric and Volatile EvolutioN SupraThermal And Thermal Ion Composition (MAVEN-STATIC) (Pickett et al., 2022). On the other hand, isotopic compositions of ions have not been detected due to the difficulty in resolving the mass difference. The Martian Moons eXploration (MMX) mission, which is planned by the Japan Aerospace Exploration Agency (JAXA) targeting the two Martian moons with the scheduled launch in 2024 (Kuramoto et al., 2022), may measure the isotopic compositions of escaping ions. It has the mass spectrum analyzer (MSA) with unprecedented mass resolution aboard MMX enough to measure isotope ratios of escaping ion species such as O$^{+}$ and C$^{+}$ (Yokota et al., 2021; Ogohara et al., 2022; Kuramoto et al., 2022). Such measurements can constrain the fractionation factor by atmospheric escape and the history of the Martian atmosphere empirically. On the other hand, it should be mentioned that if photochemical loss of carbon as a neutral is dominant then the measurements of the isotope ratios would only constrain the fractionation of a small fraction of escaping carbon.
	
\begin{figure}[htbp]
	\centering
	\includegraphics[width=0.45\columnwidth]{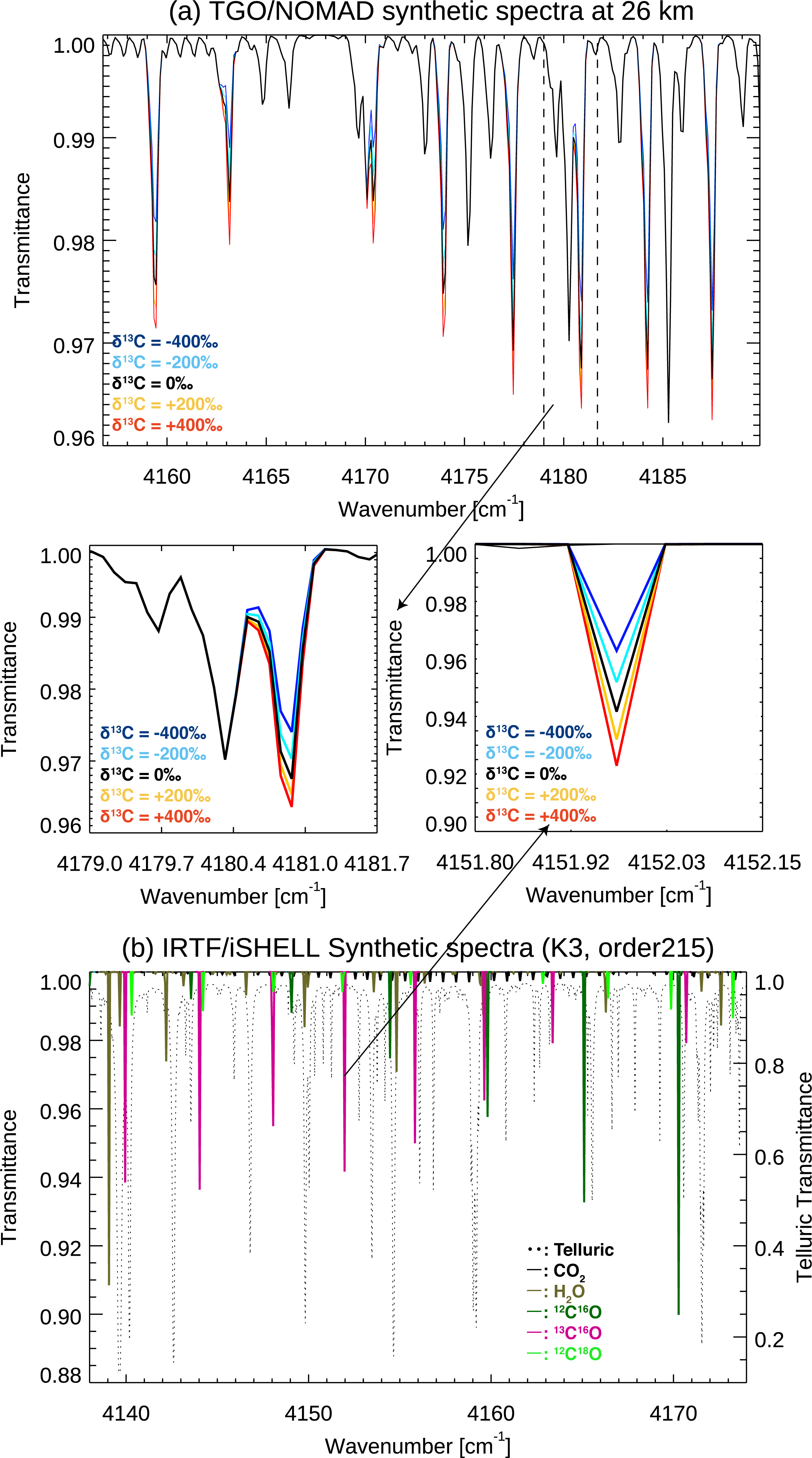}
	\caption{(a) Synthetic spectra of a solar occultation measurement by TGO/NOMAD taken with the diffraction order 186. The calculations are performed by Asimut radiative transfer and retrieval code (Vandaele et al., 2006). The simulation is made for the measurement at 26 km over the northern polar region (latitude: 82°N) at Ls=165°. The atmospheric condition is obtained from the GEM-Mars (Daerden et al., 2019). The CO volume mixing ratio at 26 km is assumed to be 776 ppm. The difference in color represents the assumed isotopic ratios for the $^{13}$CO lines (Blue: $\delta ^{13}$C=-400 ‰, Light blue: $\delta ^{13}$C=-200 ‰, Black: $\delta ^{13}$C=0 ‰, Orange: $\delta ^{13}$C=+200 ‰, Red: $\delta ^{13}$C=400 ‰). The rest of the strong features are due to $^{12}$CO. The upper panel shows the whole spectral range of the order 186, and the small bottom panel is for a limited spectral range which contains both $^{12}$CO and $^{13}$CO lines. The typical signal to noise ratio of a single NOMAD spectrum is more than 1000, which suggests that the strong depletion of $^{13}$C presented in this study may be observed by TGO/NOMAD. (b) Synthetic spectra of a measurement by IRTF/iSHELL taken with the diffraction order 215 in the K3 band. The calculations are performed with online version of Planetary Spectrum Generator (Villanueva et al., 2018). In the calculation, a typical atmospheric condition at the Mars equatorial regions (the volume mixing ratio of CO is 700 ppm) is assumed. The bottom panel shows the features due to CO$_{2}$ (the black curve), H$_{2}$O (the dark yellow curve), $^{12}$C$^{16}$O (the dark green curve), $^{13}$C$^{16}$O (the purple curve), and $^{12}$C$^{18}$O (the light green curve). The expected telluric transmittance at Maunakea observatory (where IRTF telescope is located) is exhibited as the black dotted curve. The small upper panel shows the $^{13}$C$^{16}$O lines at 4151.97 cm$^{-1}$ for different isotopic ratios (Blue: $\delta ^{13}$C=-400 ‰, Light blue: $\delta ^{13}$C=-200 ‰, Black: $\delta ^{13}$C=0 ‰, Orange: $\delta ^{13}$C=+200 ‰, Red: $\delta ^{13}$C=400 ‰). The typical signal to noise ratio of a Mars spectrum taken by IRTF/iSHELL over a few pixels (which corresponds to ~0.1”) spectra of Mars observations is more than 100, which suggests that the strong depletion of $^{13}$C presented in this study may be also observed by IRTF/iSHELL.}
\end{figure}

\section{Conclusion} \label{sec:conclusion}
	We have developed a 1-D photochemical model considering carbon isotopic fractionation induced by photolysis of CO$_{2}$ for the Martian atmosphere. According to our results, CO is depleted in $^{13}$C compared with CO$_{2}$ at each altitude due to the fractionation effect of photolysis. Below the homopause, $\delta ^{13}$C in CO decreases as the altitude decreases: it takes about $-170$ ‰ near the surface under the standard eddy diffusion setting. Above the homopouse, $\delta ^{13}$C in both CO and CO$_{2}$ decreases as the altitude increases by the diffusive separation resulting from the difference in the molecular mass between isotopologues. Our results support the hypothesis that the fractionated atmospheric CO is responsible for the production of the $^{13}$C-depleted organic carbon in Martian sediments detected by Curiosity Rover through the conversion of CO into organic materials and their deposition on the surface. The isotopic fractionation of CO by the photolysis and diffusive separation between the lower atmosphere and the escape region enhances the degree of the fractionation through photochemical escape via CO photodissociation. The fractionation factor considering these effects becomes lower than that evaluated by Hu et al. (2015): it changes from 0.60 to 0.52 under the standard eddy diffusion setting. The change in the fractionation factor may lead to a decrease in the amount lost by atmospheric escape constrained by the evolution of the atmospheric carbon isotopic composition. The fractionated isotopic composition of CO in the Martian atmosphere may be observed by ExoMars Trace Gas Orbiter (TGO) and ground-based telescopes, and escaping ion species produced by the fractionated carbon-bearing species may be detected by Martian Moons eXploration (MMX) in the future.

\appendix
\begin{table}
	\centering
	\caption{Chemical reactions}
	\label{tab:example_table}
	\begin{tabular}{llclllll} 
	\hline
	No. & Reaction &  &   & Reaction rate coefficient$^{\mathrm{a}}$ & & Column rate$^{\mathrm{b}}$\\
	& & & & & standard case & small eddy case & large eddy case\\
	\hline
	R1 & $^{12}\mathrm{CO}_{2}+h\nu$ & $\to$ & $^{12}\mathrm{CO}+\mathrm{O}$ & & $1.3\times 10^{12}$ & $1.2\times 10^{12}$ & $1.4\times 10^{12}$\\
	R2 &                                        & $\to$ & $^{12}\mathrm{CO}+\mathrm{O(^{1}D)}$ & & $2.0\times 10^{11}$ & $1.7\times 10^{11}$ & $2.1\times 10^{11}$\\
	R3 & $^{13}\mathrm{CO}_{2}+h\nu$ & $\to$ & $^{13}\mathrm{CO}+\mathrm{O}$ & & $1.2\times 10^{10}$ & $1.1\times 10^{10}$ & $1.2\times 10^{10}$\\
	R4 &                                        & $\to$ & $^{13}\mathrm{CO}+\mathrm{O(^{1}D)}$ & & $2.3\times 10^{9}$ & $1.9\times 10^{9}$ & $2.4\times 10^{9}$\\
	R5 & $\mathrm{H_{2}O}+h\nu$ & $\to$ & $\mathrm{H}+\mathrm{OH}$ & & $8.3\times 10^{9}$ & $8.5\times 10^{9}$ & $8.7\times 10^{9}$\\
	R6 &                                        & $\to$ & $\mathrm{H}_{2}+\mathrm{O(^{1}D)}$ & & $5.7\times 10^{3}$ & $5.7\times 10^{3}$ & $5.7\times 10^{3}$\\
	R7 & $\mathrm{O}_{3}+h\nu$ & $\to$ & $\mathrm{O}_{2}+\mathrm{O}$ & & $7.1\times 10^{11}$ & $5.9\times 10^{11}$ & $1.3\times 10^{11}$\\
	R8 &                                        & $\to$ & $\mathrm{O}_{2}+\mathrm{O(^{1}D)}$ & & $4.1\times 10^{12}$ & $3.4\times 10^{12}$ & $7.4\times 10^{11}$\\
	R9 & $\mathrm{O_{2}}+h\nu$ & $\to$ & $\mathrm{O}+\mathrm{O}$ & & $1.3\times 10^{11}$ & $1.3\times 10^{11}$ & $2.8\times 10^{10}$\\
	R10 &                                        & $\to$ & $\mathrm{O}+\mathrm{O(^{1}D)}$ & & $1.6\times 10^{10}$ & $5.7\times 10^{10}$ & $2.5\times 10^{9}$\\
	R11 & $\mathrm{H}_{2}+h\nu$ & $\to$ & $\mathrm{H}+\mathrm{H}$ & & $6.6\times 10^{4}$ & $6.6\times 10^{4}$ & $2.2\times 10^{5}$\\
	R12 & $\mathrm{OH}+h\nu$ & $\to$ & $\mathrm{O}+\mathrm{H}$ & & $9.0\times 10^{5}$ & $6.0\times 10^{5}$ & $1.0\times 10^{6}$\\
	R13 &                                        & $\to$ & $\mathrm{O(^{1}D)}+\mathrm{H}$ & & $4.4\times 10^{2}$ & $8.8\times 10^{1}$ & $4.8\times 10^{3}$\\
	R14 & $\mathrm{HO}_{2}+h\nu$ & $\to$ & $\mathrm{OH}+\mathrm{O}$ & & $3.9\times 10^{10}$ & $3.2\times 10^{10}$ & $3.7\times 10^{10}$\\
	R15 & $\mathrm{H_{2}O_{2}}+h\nu$ & $\to$ & $\mathrm{OH}+\mathrm{OH}$ & & $1.7\times 10^{11}$ & $1.4\times 10^{11}$ & $1.4\times 10^{11}$\\
	R16 &                                        & $\to$ & $\mathrm{HO}_{2}+\mathrm{H}$ & & $1.7\times 10^{10}$ & $1.4\times 10^{10}$ & $1.4\times 10^{10}$\\
	R17 &                                        & $\to$ & $\mathrm{H_{2}O}+\mathrm{O(^{1}D)}$ & & $0$ & $0$ & $0$\\
	R18 & $\mathrm{O}+\mathrm{O}+\mathrm{M}$ & $\to$ & $\mathrm{O_{2}}+\mathrm{M}$ & $5.4\times 10^{-33}(300/T)^{3.25}$ & $2.3\times 10^{11}$ & $2.7\times 10^{11}$ & $8.2\times 10^{10}$\\
	R19 & $\mathrm{O}+\mathrm{O_{2}}+\mathrm{^{12}CO_{2}}$ & $\to$ & $\mathrm{O_{3}}+\mathrm{^{12}CO_{2}}$ & $1.5\times 10^{-33}(300/T)^{2.4}$ & $4.9\times 10^{12}$ & $4.1\times 10^{12}$ & $9.4\times 10^{11}$\\
	R20 & $\mathrm{O}+\mathrm{O_{3}}$ & $\to$ & $\mathrm{O_{2}}+\mathrm{O_{2}}$ & $8.0\times 10^{-12}\mathrm{exp}(-2060/T)$ & $3.7\times 10^{7}$ & $4.3\times 10^{7}$ & $7.0\times 10^{6}$\\
	R21 & $\mathrm{O}+\mathrm{^{12}CO}+\mathrm{M}$ & $\to$ & $\mathrm{^{12}CO_{2}}+\mathrm{M}$ & $2.2\times 10^{-33}\mathrm{exp}(-1780/T)$ & $1.3\times 10^{8}$ & $8.9\times 10^{7}$ & $2.2\times 10^{8}$\\
	R22 & $\mathrm{O}+\mathrm{^{13}CO}+\mathrm{M}$ & $\to$ & $\mathrm{^{13}CO_{2}}+\mathrm{M}$ & $1.0074\times 2.2\times 10^{-33}\mathrm{exp}(-1780/T)$ & $1.2\times 10^{6}$ & $8.3\times 10^{5}$ & $2.1\times 10^{6}$\\
	R23 & $\mathrm{O(^{1}D)}+\mathrm{O_{2}}$ & $\to$ & $\mathrm{O}+\mathrm{O_{2}}$ & $3.2\times 10^{-11}\mathrm{exp}(70/T)$ & $2.4\times 10^{9}$ & $2.8\times 10^{9}$ & $1.0\times 10^{8}$\\
	R24 & $\mathrm{O(^{1}D)}+\mathrm{O_{3}}$ & $\to$ & $\mathrm{O_{2}}+\mathrm{O_{2}}$ & $1.2\times 10^{-10}$ & $8.4\times 10^{4}$ & $1.5\times 10^{5}$ & $2.2\times 10^{3}$\\
	R25 & $\mathrm{O(^{1}D)}+\mathrm{O_{3}}$ & $\to$ & $\mathrm{O}+\mathrm{O}+\mathrm{O_{2}}$ & $1.2\times 10^{-10}$ & $8.4\times 10^{4}$ & $1.5\times 10^{5}$ & $2.2\times 10^{3}$\\
	R26 & $\mathrm{O(^{1}D)}+\mathrm{H_{2}}$ & $\to$ & $\mathrm{H}+\mathrm{OH}$ & $1.2\times 10^{-10}$ & $5.8\times 10^{7}$ & $3.3\times 10^{7}$ & $2.5\times 10^{8}$\\
	R27 & $\mathrm{O(^{1}D)}+\mathrm{^{12}CO_{2}}$ & $\to$ & $\mathrm{O}+\mathrm{^{12}CO_{2}}$ & $7.5\times 10^{-11}\mathrm{exp}(115/T)$ & $4.3\times 10^{12}$ & $3.7\times 10^{12}$ & $9.5\times 10^{11}$\\
	R28 & $\mathrm{O(^{1}D)}+\mathrm{H_{2}O}$ & $\to$ & $\mathrm{OH}+\mathrm{OH}$ & $1.63\times 10^{-10}\mathrm{exp}(60/T)$ & $8.5\times 10^{8}$ & $5.3\times 10^{8}$ & $1.6\times 10^{8}$\\
	R29 & $\mathrm{H_{2}}+\mathrm{O}$ & $\to$ & $\mathrm{OH}+\mathrm{H}$ & $6.34\times 10^{-12}\mathrm{exp}(-4000/T)$ & $1.5\times 10^{6}$ & $1.1\times 10^{6}$ & $3.0\times 10^{7}$\\
	R30 & $\mathrm{OH}+\mathrm{H_{2}}$ & $\to$ & $\mathrm{H_{2}O}+\mathrm{H}$ & $9.01\times 10^{-13}\mathrm{exp}(-1526/T)$ & $1.3\times 10^{8}$ & $1.2\times 10^{8}$ & $1.5\times 10^{9}$\\
	R31 & $\mathrm{H}+\mathrm{H}+\mathrm{^{12}CO_{2}}$ & $\to$ & $\mathrm{H_{2}}+\mathrm{^{12}CO_{2}}$ & $1.6\times 10^{-32}(298/T)^{2.27}$ & $2.6\times 10^{5}$ & $1.9\times 10^{5}$ & $1.3\times 10^{7}$\\
	R32 & $\mathrm{H}+\mathrm{OH}+\mathrm{^{12}CO_{2}}$ & $\to$ & $\mathrm{H_{2}O}+\mathrm{^{12}CO_{2}}$ & $1.292\times 10^{-30}(300/T)^{2}$ & $1.6\times 10^{5}$ & $9.1\times 10^{4}$ & $2.0\times 10^{6}$\\
	R33 & $\mathrm{H}+\mathrm{HO_{2}}$ & $\to$ & $\mathrm{OH}+\mathrm{OH}$ & $7.2\times 10^{-11}$ & $7.0\times 10^{9}$ & $5.7\times 10^{9}$ & $6.6\times 10^{10}$\\
	R34 & $\mathrm{H}+\mathrm{HO_{2}}$ & $\to$ & $\mathrm{H_{2}O}+\mathrm{O(^{1}D)}$ & $1.6\times 10^{-12}$ & $1.6\times 10^{8}$ & $1.3\times 10^{8}$ & $1.5\times 10^{9}$\\
	R35 & $\mathrm{H}+\mathrm{HO_{2}}$ & $\to$ & $\mathrm{H_{2}}+\mathrm{O_{2}}$ & $3.45\times 10^{-12}$ & $3.4\times 10^{8}$ & $2.7\times 10^{8}$ & $3.1\times 10^{9}$\\
	R36 & $\mathrm{H}+\mathrm{H_{2}O_{2}}$ & $\to$ & $\mathrm{HO_{2}}+\mathrm{H_{2}}$ & $2.8\times 10^{-12}\mathrm{exp}(-1890/T)$ & $4.7\times 10^{5}$ & $5.0\times 10^{5}$ & $2.3\times 10^{6}$\\
	R37 & $\mathrm{H}+\mathrm{H_{2}O_{2}}$ & $\to$ & $\mathrm{H_{2}O}+\mathrm{OH}$ & $1.7\times 10^{-11}\mathrm{exp}(-1800/T)$ & $4.5\times 10^{6}$ & $4.8\times 10^{6}$ & $2.3\times 10^{7}$\\
	R38 & $\mathrm{H}+\mathrm{O_{2}}+\mathrm{M}$ & $\to$ & $\mathrm{HO_{2}}+\mathrm{M}$ & $k_{0}=8.8\times 10^{-32}(300/T)^{1.3}$ & $1.6\times 10^{12}$ & $1.4\times 10^{12}$ & $1.7\times 10^{12}$\\
	 & & & & $k_{\infty}=7.5\times 10^{-11}(300/T)^{-0.2}$\\
	R39 & $\mathrm{H}+\mathrm{O_{3}}$ & $\to$ & $\mathrm{OH}+\mathrm{O_{2}}$ & $1.4\times 10^{-10}\mathrm{exp}(-470/T)$ & $6.8\times 10^{10}$ & $8.0\times 10^{10}$ & $7.5\times 10^{10}$\\
	R40 & $\mathrm{O}+\mathrm{OH}$ & $\to$ & $\mathrm{O_{2}}+\mathrm{H}$ & $1.8\times 10^{-11}\mathrm{exp}(180/T)$ & $1.3\times 10^{11}$ & $8.7\times 10^{10}$ & $1.9\times 10^{11}$\\
	\hline
	 & & & & $^{\mathrm{a}}$2-body: cm$^{3}$s$^{-1}$; 3-body: cm$^{6}$s$^{-1}$ & & $^{\mathrm{b}}$cm$^{-2}$s$^{-1}$\\
	\end{tabular}
\end{table}

\begin{table}
	\centering
	\caption{Chemical reactions}
	\label{tab:example_table}
	\begin{tabular}{llclllll} 
	\hline
	No. & Reaction &  &   & Reaction rate coefficient$^{\mathrm{a}}$ & & Column rate$^{\mathrm{b}}$\\
	& & & & & standard case & small eddy case & large eddy case\\
	\hline
	R41 & $\mathrm{O}+\mathrm{HO_{2}}$ & $\to$ & $\mathrm{OH}+\mathrm{O_{2}}$ & $3.0\times 10^{-11}\mathrm{exp}(200/T)$ & $1.2\times 10^{12}$ & $1.1\times 10^{12}$ & $1.3\times 10^{12}$\\
	R42 & $\mathrm{O}+\mathrm{H_{2}O_{2}}$ & $\to$ & $\mathrm{OH}+\mathrm{HO_{2}}$ & $1.4\times 10^{-12}\mathrm{exp}(-2000/T)$ & $3.5\times 10^{7}$ & $3.0\times 10^{7}$ & $2.8\times 10^{7}$\\
	R43 & $\mathrm{OH}+\mathrm{OH}$ & $\to$ & $\mathrm{H_{2}O}+\mathrm{O}$ & $1.8\times 10^{-12}$ & $4.7\times 10^{5}$ & $1.6\times 10^{5}$ & $8.8\times 10^{5}$\\	
	R44 & $\mathrm{OH}+\mathrm{OH}+\mathrm{M}$ & $\to$ & $\mathrm{H_{2}O_{2}}+\mathrm{M}$ & $k_{0}=8.97\times 10^{-31}(300/T)$ & $9.5\times 10^{3}$ & $8.1\times 10^{3}$ & $9.9\times 10^{3}$\\
	 & & & & $k_{\infty}=2.6\times 10^{-11}$\\
	R45 & $\mathrm{OH}+\mathrm{O_{3}}$ & $\to$ & $\mathrm{HO_{2}}+\mathrm{O_{2}}$ & $1.7\times 10^{-12}\mathrm{exp}(-940/T)$ & $3.5\times 10^{6}$ & $3.1\times 10^{6}$ & $3.8\times 10^{5}$\\
	R46 & $\mathrm{OH}+\mathrm{HO_{2}}$ & $\to$ & $\mathrm{H_{2}O}+\mathrm{O_{2}}$ & $4.8\times 10^{-11}\mathrm{exp}(250/T)$ & $6.0\times 10^{9}$ & $5.6\times 10^{9}$ & $4.3\times 10^{9}$\\
	R47 & $\mathrm{OH}+\mathrm{H_{2}O_{2}}$ & $\to$ & $\mathrm{H_{2}O}+\mathrm{HO_{2}}$ & $1.8\times 10^{-12}$ & $2.8\times 10^{9}$ & $3.0\times 10^{9}$ & $1.5\times 10^{9}$\\	
	R48 & $\mathrm{HO_{2}}+\mathrm{O_{3}}$ & $\to$ & $\mathrm{OH}+\mathrm{O_{2}}+\mathrm{O_{2}}$ & $1.0\times 10^{-14}\mathrm{exp}(-490/T)$ & $4.2\times 10^{8}$ & $2.5\times 10^{8}$ & $7.0\times 10^{7}$\\
	R49 & $\mathrm{HO_{2}}+\mathrm{HO_{2}}$ & $\to$ & $\mathrm{H_{2}O_{2}}+\mathrm{O_{2}}$ & $3.0\times 10^{-13}\mathrm{exp}(460/T)$ & $1.8\times 10^{11}$ & $1.5\times 10^{11}$ & $1.5\times 10^{11}$\\
	R50 & $\mathrm{HO_{2}}+\mathrm{HO_{2}}+\mathrm{M}$ & $\to$ & $\mathrm{H_{2}O_{2}}+\mathrm{O_{2}}+\mathrm{M}$ & $4.2\times 10^{-33}\mathrm{exp}(920/T)$ & $6.4\times 10^{9}$ & $5.6\times 10^{9}$ & $4.9\times 10^{9}$\\
	R51 & $\mathrm{^{12}CO}+\mathrm{OH}+\mathrm{M}$ & $\to$ & $\mathrm{^{12}CO_{2}}+\mathrm{H}+\mathrm{M}$ & $k_{0}=1.5\times 10^{-13}(300/T)^{0.6}$ & $1.5\times 10^{12}$ & $1.4\times 10^{12}$ & $1.6\times 10^{12}$\\
	 & & & & $k_{\infty}=2.1\times 10^{9}(300/T)^{-6.1}$\\
	R52 & $\mathrm{^{13}CO}+\mathrm{OH}+\mathrm{M}$ & $\to$ & $\mathrm{^{13}CO_{2}}+\mathrm{H}+\mathrm{M}$ & $k_{0}=0.9891\times 1.5\times 10^{-13}(300/T)^{0.6}$ & $1.4\times 10^{10}$ & $1.3\times 10^{10}$ & $1.5\times 10^{10}$\\
	 & & & & $k_{\infty}=0.9891\times 2.1\times 10^{9}(300/T)^{-6.1}$\\
	R53 & $\mathrm{^{12}CO}+\mathrm{OH}+\mathrm{M}$ & $\to$ & $\mathrm{HO^{12}CO}+\mathrm{M}$ & $k_{0}=5.9\times 10^{-33}(300/T)^{1.4}$ & $1.1\times 10^{10}$ & $9.8\times 10^{9}$ & $9.3\times 10^{9}$\\
	 & & & & $k_{\infty}=1.1\times 10^{-12}(300/T)^{-1.3}$\\
	R54 & $\mathrm{^{13}CO}+\mathrm{OH}+\mathrm{M}$ & $\to$ & $\mathrm{HO^{13}CO}+\mathrm{M}$ & $k_{0}=0.9891\times 5.9\times 10^{-33}(300/T)^{1.4}$ & $9.9\times 10^{7}$ & $8.9\times 10^{7}$ & $8.7\times 10^{7}$\\
	 & & & & $k_{\infty}=0.9891\times 1.1\times 10^{-12}(300/T)^{-1.3}$\\	 	 
	R55 & $\mathrm{HO^{12}CO}+\mathrm{O_{2}}$ & $\to$ & $\mathrm{HO_{2}}+\mathrm{^{12}CO_{2}}$ & $2.0\times 10^{-12}$ & $1.1\times 10^{10}$ & $9.8\times 10^{9}$ & $9.3\times 10^{9}$\\
	R56 & $\mathrm{HO^{13}CO}+\mathrm{O_{2}}$ & $\to$ & $\mathrm{HO_{2}}+\mathrm{^{13}CO_{2}}$ & $0.9891\times 2.0\times 10^{-12}$ & $9.9\times 10^{7}$ & $8.9\times 10^{7}$ & $8.7\times 10^{7a}$\\
	R57 & $\mathrm{^{12}CO_{2}^{+}}+\mathrm{H_{2}}$ & $\to$ & $\mathrm{^{12}CO_{2}}+\mathrm{H}+\mathrm{H}$ & $8.7\times 10^{-10}$ & $1.4\times 10^{8}$ & $1.2\times 10^{8}$ & $1.4\times 10^{9}$\\
	\hline
	 & & & & $^{\mathrm{a}}$2-body: cm$^{3}$s$^{-1}$; 3-body: cm$^{6}$s$^{-1}$ & & $^{\mathrm{b}}$cm$^{-2}$s$^{-1}$\\
	\end{tabular}
\end{table}



\end{document}